\documentclass[3p]{elsarticle}
\usepackage{amsmath}
\usepackage{mathtools}
\usepackage{amssymb}
\usepackage{graphicx}
\usepackage{amsbsy}
\usepackage{caption}
\usepackage{subcaption}
\usepackage{lineno,hyperref}
\usepackage{float}
\usepackage{multirow}
\usepackage{color}
\modulolinenumbers[5]
\usepackage{tabularx}
\usepackage[toc,page]{appendix}
\biboptions{numbers,sort&compress}
\DeclareMathOperator*{\argmax}{argmax}
\usepackage{nomencl}
\makenomenclature

\journal{Buildings and Environments}

\begin{document}


\begin{frontmatter}
\title{A Transfer Operator Methodology for Optimal Sensor Placement Accounting for Uncertainty}

\author[mymainaddress]{Himanshu Sharma\corref{}}
\cortext[mycorrespondingauthor]{Corresponding author}
\author[mysecondaryaddress]{Umesh Vaidya\corref{}}
\author[mymainaddress]{Baskar Ganapathysubramanian\corref{mycorrespondingauthor}}

\ead{baskarg@iastate.edu}

\address[mymainaddress]{Department of Mechanical Engineering, Iowa State University, Ames, IA 50010, USA}
\address[mysecondaryaddress]{Department of Electrical and Computer Engineering, Iowa State University,Ames,IA,50010,USA}


\begin{abstract}
Sensors in buildings are used for a wide variety of applications such as monitoring air quality, contaminants, indoor temperature, and relative humidity. These are used for accessing and ensuring indoor air quality, and also for ensuring safety in the event of chemical and biological attacks. It follows that optimal placement of sensors become important to accurately monitor contaminant levels in the indoor environment. 
However, contaminant transport inside the indoor environment is governed by the indoor flow conditions which are affected by various uncertainties associated with the building systems including occupancy and boundary fluxes. Therefore, it is important to account for all associated uncertainties while designing the sensor layout. 
The transfer operator based framework provides an effective way to identify optimal placement of sensors. Previous work has been limited to sensor placements under deterministic scenarios. In this work we extend the transfer operator based approach for optimal sensor placement while accounting for building systems uncertainties. The methodology provides a probabilistic metric to gauge coverage under uncertain conditions. We illustrate the capabilities of the framework with examples exhibiting boundary flux uncertainty. 
\end{abstract}
\end{frontmatter}
\section{Introduction}
With increasing miniaturization in electronic components the deployment of multiple sensors is becoming an integral part of daily life. These sensors are typically interconnected (either locally or centrally), and make up a sensor network. The spatial placement of sensors that make up the network is especially important in the context of ensuring indoor air quality (IAQ) in the built environment. This is because an individual in the western world spends about 90\% of their lifetime indoor (World Health Organization~\citep{WHOR1}). Degraded IAQ is linked to various respiratory diseases \citep{Tham2016}, with long term exposure linked to serious illness and death~\citep{herbstman2015prenatal,baur1993health}. Additionally, in large public spaces such as atriums, airports, subway stations, and indoor stadiums the assessment of indoor air quality is critical from the security and safety standpoint. This is especially important to avoid transmission of infectious diseases (TID) such as influenza, tuberculosis, and SARS. The spread of such diseases in the built environment has been widely studied \citep{Olsen2003,Namilae2017,Wenzel1996,Kawashima2016,Li2016,Buhr2016} and is closely connected with air flow and contaminant/infection  transport. An optimally placed set of sensors can effectively mitigate these problems. \\
\indent For designing sensor networks to monitor IAQ in indoor spaces, a complete knowledge of air-flow distribution is needed. Additionally, the sensor network must satisfy several criteria such as coverage, sensitivity, response time and detection threshold~\citep{Liu2009}. Broadly, there are three approaches for sensor placement in the built environment: 1) engineering and heuristic methods, 2) optimization and inverse methods, and 3) dynamical systems based methods. The engineering methods are mostly thumb rule based. They usually do not take into account the flow in the space. This affects the total coverage of the sensor network and leaves some regions uncovered. While very simple to use, this approach results in expensive sensor networks, as more number of sensor are required to improve the coverage of the space~\citep{Liu2009}. The optimization and inverse methods were developed to overcome the shortcomings of the heuristic approach. These methods use the actual flow fields or flow rates -- either from CFD simulations or from zonal models. A cost function is constructed that encodes the coverage of sensing along with additional constraints such as cost, and detection threshold. This cost function is then minimized to identify the optimal sensor locations~\citep{Eliades2013,Liu2009,Zhang2007,Chen2008,Chen2010}.\\
\indent The optimization based techniques has been successfully used for a wide variety of sensor placement scenarios~\citep{Mazumdar2008,arvelo2002enhanced,Chen2012,Zhai,Chen2010} using both CFD or zonal simulations. While these methods account for the effects of flow fields in the sensor placement process, challenges include computational difficulties in finding the optima. Furthermore, these methods rely on solving the advection-diffusion equation for multiple release scenarios and are based on iterative solves which makes them very computationally challenging. Similar issues exist with the inverse and adjoint based methods~\citep{Liu2009,Liu2009a}. The choice of the initial guess is extremely crucial in these approaches, and typically work well when the number of sensors to be placed is small. Finally, this approach can only find locally optimal sensor locations, with no guarantees of global optima.\\
\indent The third category of methods is based on a dynamical systems approach to identify the sensor placement locations~\citep{sinha2016operator,sinha2013optimal,vaidya2012actuator}. The approach is based on the construction of the Perron-Frobenius (PF) operator to determine the advection of a contaminant subjected to a non-linear flow field. The key advantage behind the PF operator framework is that the nonlinear evolution of contaminant under nonlinear and uncertain flow fields can be replaced by a equivalent (but high-dimensional) linear evolution of contaminant density. The linear nature of the PF operator affords three significant capabilities: First, this allows well developed concepts from linear control theory to be applied to contaminant transport analysis, second, the resulting algorithms are based on simple matrix-vector products, which make implementation and deployment extremely simple and efficient (both in terms of memory as well as computation), and third, the approach provides guarantees of finding the globally optimal sensor locations. Fontanini et. al~\citep{Fontanini2015,Fontanini2017} showed the construction and usage of the discrete PF operator for fast computation of contaminant transport in the built environment. They then used the PF operator to identify optimal sensor locations under deterministic flow scenarios~\citep{Fontanini2016}. \\
\indent In recent years there has been increasing interest in incorporating the effects of various uncertainties in the building during analysis of building performance \citep{hopfe2013multi,hopfe2011uncertainty,sun2014exploring}. The uncertainties associated with buildings can be broadly classified into three categories: (1) environmental, (2) workmanship and quality of building elements, and (3) behavioural. Environmental factors consist of location-specific weather conditions which a building is subjected to. This appears as uncertain boundary conditions. Workmanship and quality is associated with the quality and design of building and construction materials. This appears as uncertainty in material properties used in the analysis. Finally, the uncertainties associated with human behaviour, such as occupancy, interior design and usage of appliances are categorized under behavioural uncertainty. Previous studies have shown the importance of these uncertainty on the energy demands \citep{yan2017quantifying} as well as on the indoor air quality \citep{zhu2007uncertainty}. Such prior work has emphasized the importance of accounting for the uncertainty associated with building systems, and motivates this study to develop a sensor placement strategy which can account for uncertainties. \\
\indent In the present work we extend the dynamical systems based approach developed by Fontanini et.al \citep{Fontanini2016, fontanini2016quantifying} to account for various building uncertainties while performing sensor placements. We specifically focus on uncertainties associated with environmental effects as they have been shown to significantly affect the flow dynamics within the built environment\footnote{The occupancy affects the flow dynamics, and we have previously shown how to account for this uncertainty~\cite{sharma2018transfer}. The quality and workmanship does plays a relatively smaller role due to the fact that there are well defined standards (like ASTM,ASHRAE) that minimize variability}. Our main contributions are as follows: (a) we contribute to the development of robust sensor placement methods by developing a novel optimal sensor placement algorithm which can account for environmental  uncertainties and can be easily extended to account for other uncertainties like occupant behavioural uncertainties, (b) we show implementation of the approach for actual complex building geometries, and (c) show incorporation of sensing constraints such as sensor location, sensing area and accounting for sensor quality to plan the sensor network.\\
\indent The outline of the paper is as follows: We first briefly outline the deterministic approach in Section~\ref{sec:MethodD} before extending the formulation to include uncertain effects in Section~\ref{sec:MethodS}. In Section~\ref{sec:results} we illustrate the method with representative examples using 2D and 3D complex geometries. Finally, we detail conclusions and future avenues of work in Section~\ref{sec:Conclude}.

\section{Brief overview of the deterministic PF approach}\label{sec:MethodD}
\subsection{Problem definition}\label{subsec:probdef}
Consider a domain $\Omega \subset \mathbb{R}^{d}$ and a given non-linear (un)steady flow field, $\mathbf{U}$, in the domain $\Omega$. Suppose a contaminant is released in the domain, the aim is to find the sensor locations in $\Omega$ that will optimally detect the contaminant release, within the response time of sensors. The sensor placement problem can be posed in a few different ways, depending on what constraints are important to respect. Different constraints result in the following scenarios for sensor placement: (a) Placement when the number of sensors is fixed; (b) Placement when a threshold coverage is desired, with no limit on the number of sensors; and (c) Placement when there are sensing and placement constraints. The third scenario can be further divided into cases where (i) the sensor locations are constrained to be placed outside the occupied zone, (ii) the sensor coverage is only needed in the occupied zone, and (iii) sensor locations are constrained to be placed outside the occupied space, and sensor coverage is needed only for the occupied space. We briefly review the PF based approach developed in Fontanini et.al \citep{Fontanini2016} that naturally accounts for all these scenarios.

\subsection{A brief overview of the deterministic approach for PF based sensor placement} \label{subsec:determinsticapprch} 
The dynamical systems approach ~\citep{Rajaram2010,vaidya2012actuator} utilizes the Perron-Frobenious (PF) operator for modeling the evolution of contaminants under non-linear vector fields. The contaminant history is then used to formulated the problem of sensor placement by accounting for various constraints. We first briefly describe how the PF operator is constructed, followed by the deterministic sensor placement approach. Following this, we extend to the uncertain case.  

\subsubsection{Contaminant Transport and Construction of the PF Operator}\label{subsec:ConsPFOpertor}

\begin{figure}
\centering
\includegraphics[scale=0.25]{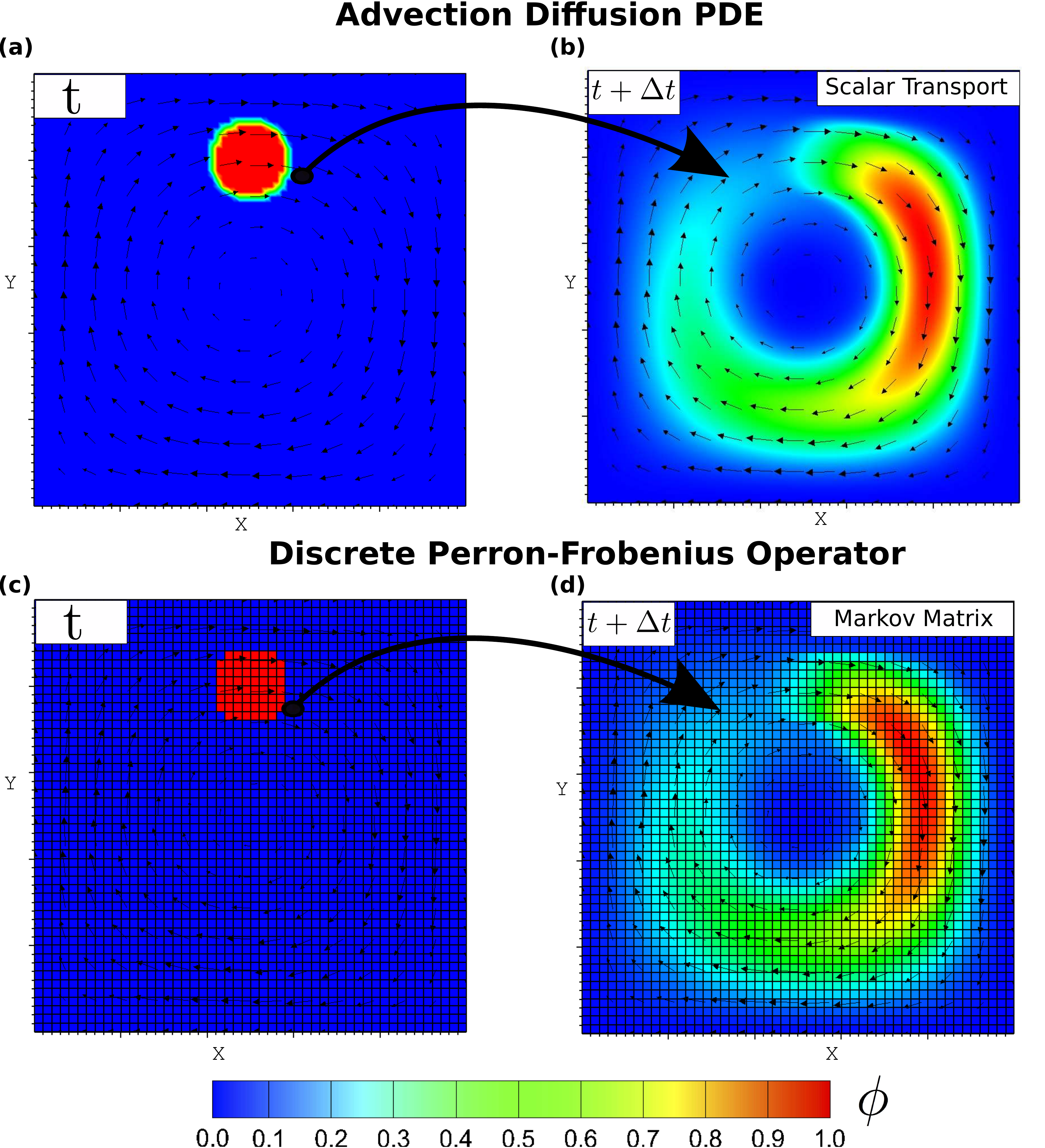}
\caption{For the given velocity field (a)-(b)Shows the contaminant transport using the scalar transport advection diffusion equation-\ref{Eq:scalarTran}(c)-(d) Shows the discrete PF-operator based scalar transport \citep{Fontanini2016}}.
\label{Fig:Euler_transp}
\vspace*{-1em}
\end{figure}

Contaminant transport is modelled by the advection-diffusion partial differential equation given by Eq-\ref{Eq:scalarTran}. \begin{equation}
\frac{\partial \Phi}{\partial t} + \nabla(U\Phi) +\nabla^{2}(D\Phi) = S_{\Phi}
\label{Eq:scalarTran}
\end{equation}
The contaminant density, $\Phi$, is propagated by the air flow field, $U$. The flow field can be generated either experimentally, or computationally using computational fluid dynamics (CFD). Here $D$ is the diffusion constant and $S_{\Phi}$ is the source term. The velocity flow field, $U$, can be steady or can be a function of time. Eq.\ref{Eq:scalarTran} can be thought of as an operator (acting on the field $\Phi$) transporting contaminant from time $t$ to $t+\delta t$~ \citep{Vaidya2012,Fontanini2015,Fontanini2016}. This operator is called the Perron-Frobenius (PF) operator $L(\cdot)$ Eq.(\ref{Eq:2}). 
\begin{equation}
\Phi_{t+\delta t} = L(\Phi_{t})
\label{Eq:2}
\end{equation}
The discrete form of this operator is called the Markov matrix $\mathbf{P}$. The Markov matrix is a non-negative square matrix with several useful properties including row-, and column- stochasticity (i.e. row- and column- elements sum to 1) and unit eigen spectrum. There are several approaches to calculate the entries of the Markov matrix, especially for contaminant transport~\citep{Chen2015,Fontanini2015,Fontanini2017}. We specifically use the Eulerian based method \cite{Fontanini2017}, as this places no restrictions on the Markov time step~\citep{Chen2015,Fontanini2015}.\\
Once constructed, the Markov matrix can be used to propagate the concentration field at discrete time instances, $t_{i}$, to reach a time horizon $t_{f} = t + m \delta t$. This propagation is performed by simple matrix-vector multiplication \footnote{We abuse notation to interchangeably refer to $\Phi$ as a field, and as its equivalent vector form}, Eq.(\ref{Eq:3}). 
\begin{equation}
\Phi_{t_{i+1}} = \Phi_{t_{i}} \mathbf{P} + \hat{S}_{t_{i},t_{i+1}} \hspace{2 mm} i \in \{0,\dots,m \} 
\label{Eq:3}
\end{equation}
The source term, $\hat{S}_{t_i,t_{i+1}}$, includes volumetric and inlet sources in the domain \citep{Fontanini2017}. An example demonstrating the use of this operator is shown in Fig.\ref{Fig:Euler_transp}. 

The next step after constructing the $\mathbf{P}$ matrix is to calculate the contaminant tracking matrix $\mathbf{Q}_{\tau}$.

\subsubsection{Contaminant tracking matrix}
\begin{figure}[htp]
\centering
\includegraphics[scale=0.9]{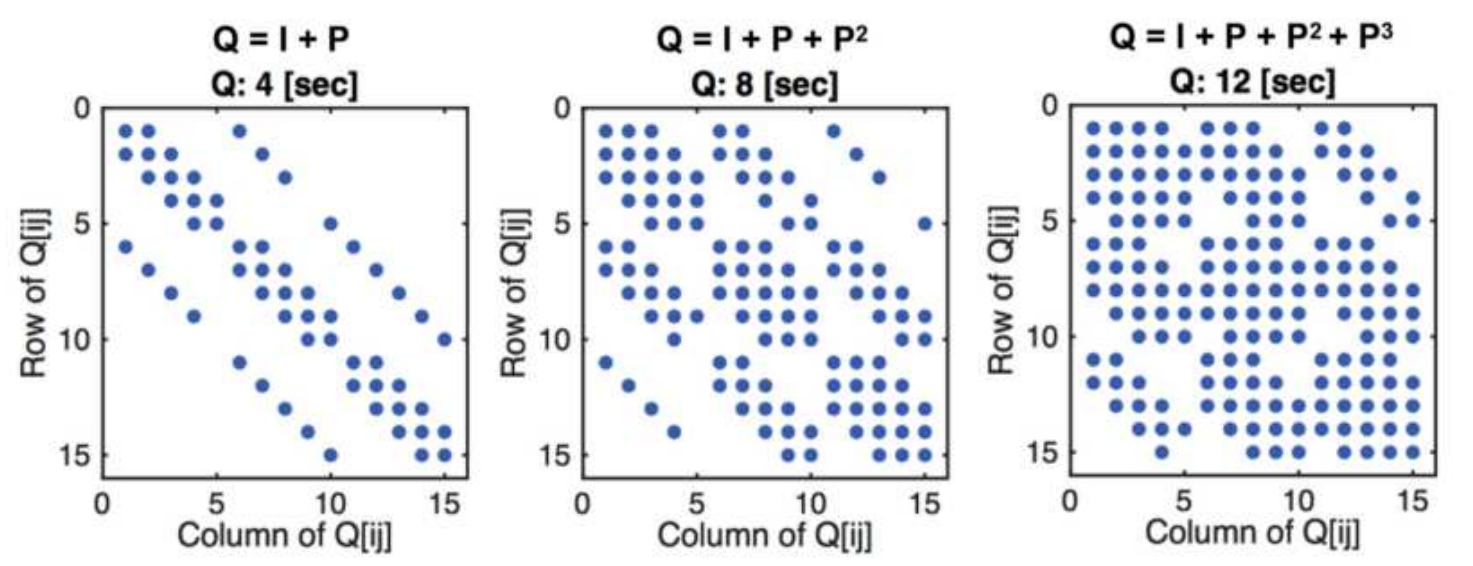}
\caption{The contaminant tracking matrix calculated from Eq.(\ref{Eq:contamTran}) for 4,8 and 12 sec of an aircraft cabin \cite{Chen2014}}
\label{Fig:Chen Qmatrices}
\end{figure}

 The Markov matrix entries correspond to the probabilities of moving from one state to another in a time step $\delta t$. The next step is to use this matrix to construct the contaminant tracking matrix $\mathbf{Q}_{\tau}$, to find how a constant contaminant source is transported over time interval $\tau = m\times \delta t$. This can can be done as follows:
\begin{equation}
\mathbf{Q}_{\tau} =  \mathbf{I} + \mathbf{P} + \mathbf{P}^2 + \mathbf{P}^3 + \dots + \mathbf{P}^m 
\label{Eq:contamTran}    
\end{equation}
$\mathbf{Q}_{\tau}$ is the matrix which encodes the contaminant tracking history, i.e. where the contaminant will propagate in time $\tau$. Figure-\ref{Fig:Chen Qmatrices} shows how we construct the contaminant transport for a representative problem of airflow in an aircraft cabin studied by Chen et. al \cite{Chen2014}.

Once the contaminant tracking matrix is constructed the sensor locations can be decided using $\mathbf{Q}_{\tau}$. 

\subsubsection{Sensor placement}
The contaminant tracking matrix, $\mathbf{Q}_{\tau}$, encodes all release scenarios. Each row, $i$, of the matrix encodes a release scenario, with contaminant release from state $i$. The non-zero elements of that row indicate which states contain contaminant that was released from state $i$. It was shown by Vaidya et. al. \citep{vaidya2012actuator} that the optimal states to place sensors are those that are most observable, i.e. the columns of $\mathbf{Q}_{\tau}$ that are most non-zero. Choosing the sensor locations becomes the so-called \textit{set cover problem} of combinatorics~\citep{Fontanini2016}. This essentially identifies a set of locations that maximize the coverage of states. A \textit{greedy algorithm} \citep{chvatal1979greedy} is used to find the sensor locations according to the maximum number of covered states.\\

\subsubsection{Applying sensing constraints} \label{subsec:SensingConstraint}
There are various constraints associated with sensors that have to be accounted for while deciding on the sensor location. Each sensor has an associated measurement accuracy threshold, which depends on the type/quality of the sensor. The accuracy threshold can be accounted for by inspecting the column entries of the contaminant transport matrix, $\mathbf{Q}_{\tau}$. The entries with larger values represent a stronger signal. The objective of this step is to replace those entries from the contaminant transport matrix which are below a sensing threshold. The threshold value is a non-dimensional sensing threshold that is the ratio of value detected to the product of the source release rate $S_{source}$ and the release time $\tau$, and is defined as
$ \epsilon_{acc} = \frac{\mu_{detect}}{S_{source} \tau} = \frac{\mu_{detect}}{\mu_{source}}$. The values $\epsilon_{acc}$ are usually prescribed by the sensor manufacturer. Since we have constructed the matrix ${\mathbf{Q}}_{\tau}$ we can apply this threshold on the matrix as Eq.(\ref{Eq:5}).
\begin{equation}
{\mathbf{Q}^{*}_{\tau}} = 
\begin{dcases} 1 & \text{if}~{\mathbf{Q}_{\tau}} \geq \epsilon_{acc} \\
0,              & \text{otherwise}
\end{dcases}
\label{Eq:5}
\end{equation}
This equation serves as an operator that converts a matrix with real entries, $\mathbf{Q}_{\tau}$, to a matrix with binary entries, $\mathbf{Q}^{*}_{\tau}$. The entries in $\mathbf{Q}^{*}_{\tau}$ that are 1 correspond to the states that can be sensed by sensors with accuracy of $\epsilon_{acc}$ if a sensor is located in the state corresponding to the column in which the entry resides.\\
\indent Analogously, not every location in the domain is suitable to place a sensor due to practical reasons such as aesthetics, sensor installation limitations, and operation requirements (occupants can affect the sensor operation by stepping over them). Therefore, it is important to account for constraints associated with sensor placement selection. In a broad sense these constraints can classified as:  1) location constraint, where the sensors must be placed outside the occupied zone of the building; and 2) sensing constraint, where we are interested in {\emph{only}} the occupied zone. To account for the location constraint, if $(\mathbb{N}^{nloc})$ states in the domain are not able to accommodate a sensor then the columns corresponding to those states are removed and replaced by zeros: 
\begin{equation}
    \mathbf{Q}^{*}_{\tau}(:,j) = 0 ; \qquad j \in \mathbb{N}^{nloc} 
    \label{Eq:SenConstraints}
\end{equation}
Similarly, for accounting for the sensing constraints, the states $(\mathbb{N}^{nsen})$ which are not in the occupied space are unimportant, and are replaced by zeros:
\begin{equation}
    \mathbf{Q}^{*}_{\tau}(i,:) = 0 ; \qquad i \in \mathbb {N}^{nsen}
    \label{Eq:LocConstraints}
\end{equation}


\section{Extension to the uncertain case}\label{sec:MethodS}
We have detailed the PF based methodology for sensor placement under a deterministic flow field. A building in general can be subjected to various weather scenarios leading to different HVAC conditions, interior arrangements and occupancy conditions. These changes brings uncertainty in the velocity field (airflow) inside the building. Therefore, it is important to account this distribution of velocity fields that can occur to identify the sensor locations which can perform optimally in every flow condition. Wang et.al \citep{wang2012uncertainties} showed that HVAC performance is significantly affected by weather conditions. The HVAC operation and weather conditions together significantly affect the flow dynamics in the building. Therefore, in the current study we illustrate the sensor placement approach by accounting for weather uncertainty. 

\subsection{Problem definition under uncertainty}
Consider the domain $\Omega \subset \mathbb{R}^d$ with the boundary $\partial \Omega$. 
We consider a set $\widetilde{\pmb{\xi}}(t) \in \mathbb{R}^n$ of random variables, where $n$ is the number of random variables which affect the flow field \footnote{We assume that all random variables are independent identically distributed (i.i.d.). Note that at places we ignore the temporal dependence of the random variable $\widetilde{\pmb{\xi}}$ for notational convenience.}
As discussed earlier, the flow field is affected by the boundary conditions (environment) of the domain, occupancy and material properties. These uncertainties can be represented in terms of a set of random variables, $\widetilde{\pmb{\xi}}$~\cite{guo2016constructing,Fontanini2013,Baskar2008b,Baskar2008a}. 
The flow field in the domain is now a function of these random variables, $\mathbf{U} \equiv \mathbf{U}(\widetilde{\pmb{\xi}}(t))$. This definition naturally allows extension to account for other uncertain parameters.  With $\mathbf{U}(\widetilde{\pmb{\xi}}(t))$, stochastic, the advection-diffusion equation (\ref{Eq:scalarTran}) transforms to a stochastic partial differential equation. 

The problem definition now becomes: find sensor locations that optimally sense contaminant distributions under uncertain flow conditions.  In contrast to the deterministic case that maximizes the coverage using the PF operator for the deterministic flow field, we will use PF operator constructed for the stochastic flow field to maximize the coverage. We show that this approach essentially lead to the {\it maximization of the expectation of the coverage}.

The construction of PF operator for the case when the velocity field, $\mathbf{U}(\widetilde{\pmb{\xi}})$, is stochastic will proceed as follows: We sample the stochastic velocity field and create a finite set of possible realizations with associated probabilities of occurrence. We then construct the PF operators for each of these realizations, and construct the contaminant transport matrix, $\mathbf{Q}_{\tau}$. We then evaluate the expected coverage using the set of contaminant transport matrices and their associated probabilities. We detail each of these steps next. 

\subsection{Constructing PF operators and Contaminant Transport for the sampling points}\label{subsub:SamplingandCons}
\indent We construct a finite set of realizations of a random variable by choosing $M$ samples from the distribution of $\widetilde{\pmb{\xi}}$. The $M$ samples forming the set $\mathbf{S} = \{s_{1},\dots,s_{M}\}$ are chosen such that they represent the $\widetilde{\pmb{\xi}}$ distribution in a statistical sense. Specifically, associated with each sample is a probability of occurrence (weight) represented as $\pmb{\Theta}=\{\theta_{1},\dots,\theta_{M}\}$ satisfying, $\theta_i\geq 0$ and  $\sum_i \theta_i=1$. Thus, each sample, $s_i$ has an associated flow field, $\mathbf{U}(s_i)$, along with a probability of occurrence of this flow field. 

The set, $(\mathbf{U}(\mathbf{S}), \pmb{\Theta})$ is chosen such that they produce the same expectation as the full distribution,
\begin{equation}
    \int \mathbf{U}(\widetilde{\pmb{\xi}}) d{\widetilde{\pmb{\xi}}} = \sum_{i=1}^{M}\theta_i*\mathbf{U}(s_i)
\end{equation}
Given the probability distribution of the random variable there are standard approaches to choosing the sampling points and associated weights \cite{mckay2000comparison,huntington1998improvements,spanier1994quasi,tokdar2010importance}. These are discussed in detail in appendix-\ref{appen}.

\indent In the current work we use CFD simulations to calculate the flow field corresponding to each sample point. In our case, the sample points correspond to realizations of different boundary conditions arising from weather variability that affect the flow field. Therefore, for each sample point a flow field is computed which is collectively represented as $\widetilde{\mathbf{U}} = \{\mathbf{U}_{1},\dots,\mathbf{U}_{M}\}$. Now, using the approach discussed in section-\ref{subsec:ConsPFOpertor} we construct the PF operators $\widetilde{\mathbf{P}} = \{ \mathbf{P}_{1},\dots,\mathbf{P}_{M} \}$ corresponding to each velocity field in $\widetilde{\mathbf{U}}$. Under the i.i.d. assumption on the random variable ${\widetilde{\pmb{\xi}}}(t)$, the stochastic counterpart of the deterministic PF operator (Eq.~\ref{Eq:3}) for the propagation of contaminant can be written as \cite{froyland2001extracting}
\begin{equation}
\Phi_{t_{j+1}} = \Phi_{t_{j}} \sum_{i=1}^M \theta_i \mathbf{P}_i + \hat{S}_{t_{j},t_{j+1}} \hspace{2 mm} j \in \{0,\dots,m \} 
\label{Eq:stochasticPF}
\end{equation}
This equation indicates that the PF operator for the stochastic flow field is just the probability weighted sum of the individual PF operators i.e., the expected PF operator. Now due to the {\it linear nature of the proposed transfer operator framework, the use of expected PF operator,  $\sum_{i=1}^M \theta_i \mathbf{P}_i$, to maximize the coverage will essentially lead to maximization of the expectation of the coverage}.

Next, the contaminant transport matrices $\widetilde{\mathbf{Q}}=\{ \mathbf{Q}_{\tau,1},\dots,\mathbf{Q}_{\tau,M}\}$ for each element of the set $\widetilde{\mathbf{U}}$ are constructed. The final step is to threshold the set of matrices based on sensor accuracy, and apply any location and sensing constraints. The procedure discussed in section-\ref{subsec:SensingConstraint} can be applied to each contaminant tracking matrix by applying Eqs.~\ref{Eq:5},\ref{Eq:SenConstraints},\ref{Eq:LocConstraints}, resulting in $ \widetilde{\mathbf{Q}^*_{\tau}}=\{\mathbf{Q}^*_{\tau,1},\dots,\mathbf{Q}^*_{\tau,M}\}$. 

\subsection{Calculation Expected Volumetric Coverage and Sensor locations}
Coverage for a sensor network is interpreted as how well the sensors can monitor the domain. This is quantified in terms of what fraction of the volume of the domain is being sensed. Considering the fact that the CFD simulations are usually performed on a nonuniform discretization of the domain, we scale the value of each state in $\mathbf{Q}^*_{\tau,i}$ with the volume of that state. Thus $\mathbf{Q}^{**}$ is the volumetric scaled version of $\mathbf{Q}^{*}$, and each row of it is computed as
\begin{equation}
    \mathbf{Q}^{**}(i,:) = \mathbf{Q}^{*}(i,:)*V_{\omega_i}/ V_{\Omega tot}
\end{equation}
where $V_{\omega_i}$ is the volume of each state and $V_{\Omega tot}$ total volume of domain $\Omega$. We compute the volumetrically scaled version of the coverage for each contaminant tracking matrix to get $\widetilde{\mathbf{Q}^{**}} =  \{\widetilde{\mathbf{Q}^{**}}_{1},\cdots, \widetilde{\mathbf{Q}^{**}}_{M}\}$. Next, we calculate the total coverage produced if a sensor is placed in state $j$, for every state $j = 1,\cdots, N$ in the domain. This is constructed simply by computing the column sum of the $\widetilde{\mathbf{Q}^{**}}$ matrices
\begin{equation}
  \mathbf{\bar{v}}(j) = \sum_{k=1}^N \widetilde{\mathbf{Q}^{**}}(j,k)   
\end{equation}
This operation is mathematically equivalent to the column wise $L_1$ norm of the matrix, and results in a vector $\mathbf{\bar{v}}$ of size $N \times 1$. We perform this operation to get the coverage for each of the $M$ samples to get
\begin{equation}
    \mathbb{V} = \{\mathbf{\bar{v}}_1, 
    \dots,\mathbf{\bar{v}}_M\}
    \label{Eq:coveragecompute}
\end{equation}

Finally, the \textbf{\textit{expected coverage}} is evaluated as the weighted sum of the coverage of each sample, $\bar{\mathbf{v}}_{M}$, with the associated probabilities $\theta_{M}$. The expected coverage, $\mathbb{E}[{\mathbb{V}}]$, is obtained as
\begin{equation}
\mathbb{E}[{\mathbb{V}}] = \sum_{i=1}^{M} \bar{\mathbf{v}_{i}}\theta_{i} 
\label{Eq:Expectaionscompute}
\end{equation}
Once we have calculated the expected coverage vector, $\mathbb{E}[{\mathbb{V}}]$, the location of the first sensor,  $\bar{k}(1)$, is the state (i.e. element of the expected coverage vector), $k$ of $\mathbb{E}[{\mathbb{V}}]$ with maximum value:
\begin{equation}
\bar{k}(1) = \argmax_k \mathbb{E}[\bar{\mathbb{V}}](k)
\label{Eq:PlaceSens}
\end{equation}
As in the deterministic case, the identification of the next sensor is done iteratively after removing the coverage of the previously placed sensors~\cite{Fontanini2016}.  
To find the next sensor, the information of the first sensor is removed from consideration (since we want to ensure that the next sensor has coverage over parts of the domain that are not covered by the first sensor). This is done by zeroing out the row and column corresponding to the state, $\bar{k}(1)$, for each  $(\widetilde{\mathbf{Q}^{**}})_{i}$ matrix. Then the sensor placement procedure is repeated ( i.e. Eq.\ref{Eq:coveragecompute}-\ref{Eq:PlaceSens}) until all the sensors are placed in the domain or the required criteria for coverage is full filled. 

\subsection{Overview of algorithm}
\begin{figure}[hp]
\centering
\includegraphics[scale=0.55]{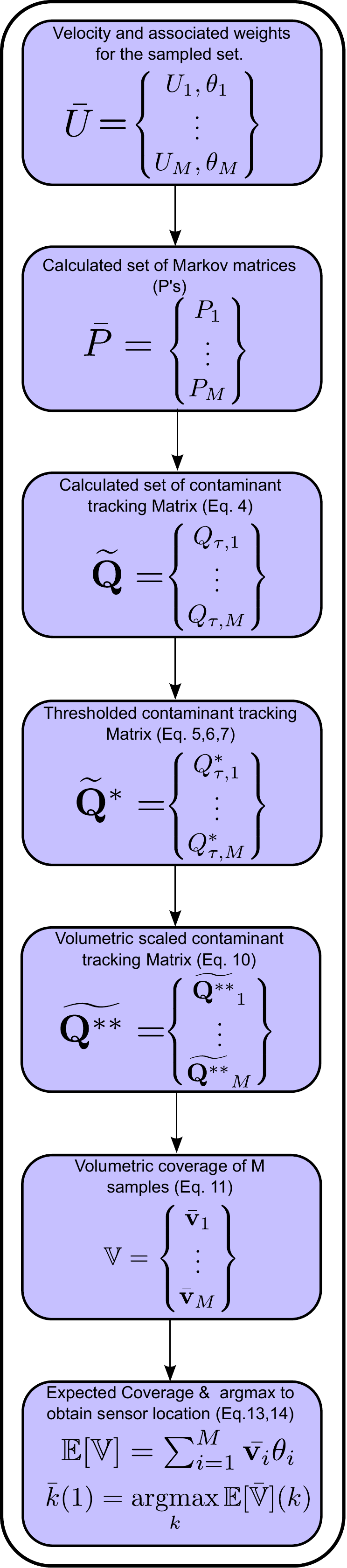}
\caption{The flow chart of the sensor placement algorithm to account for the uncertain flow fields inside the buildings.}
\label{Fig:uncertainflowchart}
\end{figure}

 The complete expectation based algorithm is illustrated as a flow chart in~Fig-\ref{Fig:uncertainflowchart}. We start with a finite set of velocity realizations and associated probability weights (with a detailed discussion of sample choices in appendix~A). Step~2: For each sample realization, the corresponding PF (or Markov) matrix, $\mathbf{P}_{i}$, is constructed from the velocity field $\mathbf{U}_{i}$ for a specified time horizon ($\tau$). Step~3: The set of contaminant tracking matrices, $\mathbf{Q}_{\tau,i}$ are constructed. Step~4: Operational scenarios are accounted for by applying sensor accuracy thresholding and constraints. This results in the set of matrices, $\mathbf{Q}^{*}_{\tau,i}$ Step~5: Each state is weighted by its volume to get the volumetrically scaled contaminant tracking matrices,$\mathbf{Q}^{**}_{i}$ . Step~6: The volumetric coverage, $\bar{\mathbf{v}_{i}}$, of each matrix is computed. Step~6: The expected coverage $\mathbb{E}[{\mathbb{V}}]$ is computed and the sensor locations calculated by finding the index of the maximum entry in the expected vector. 
 
\section{Results and Discussions}\label{sec:results}
This section presents results of the algorithm for sensor placement under uncertain operating conditions for a set of benchmark buildings.
\subsection{Problem geometry and uncertain boundary conditions}\label{subsec:ExampleProb}

\begin{figure}[htp]
\centering
\includegraphics[scale=0.5,width=0.75\linewidth]{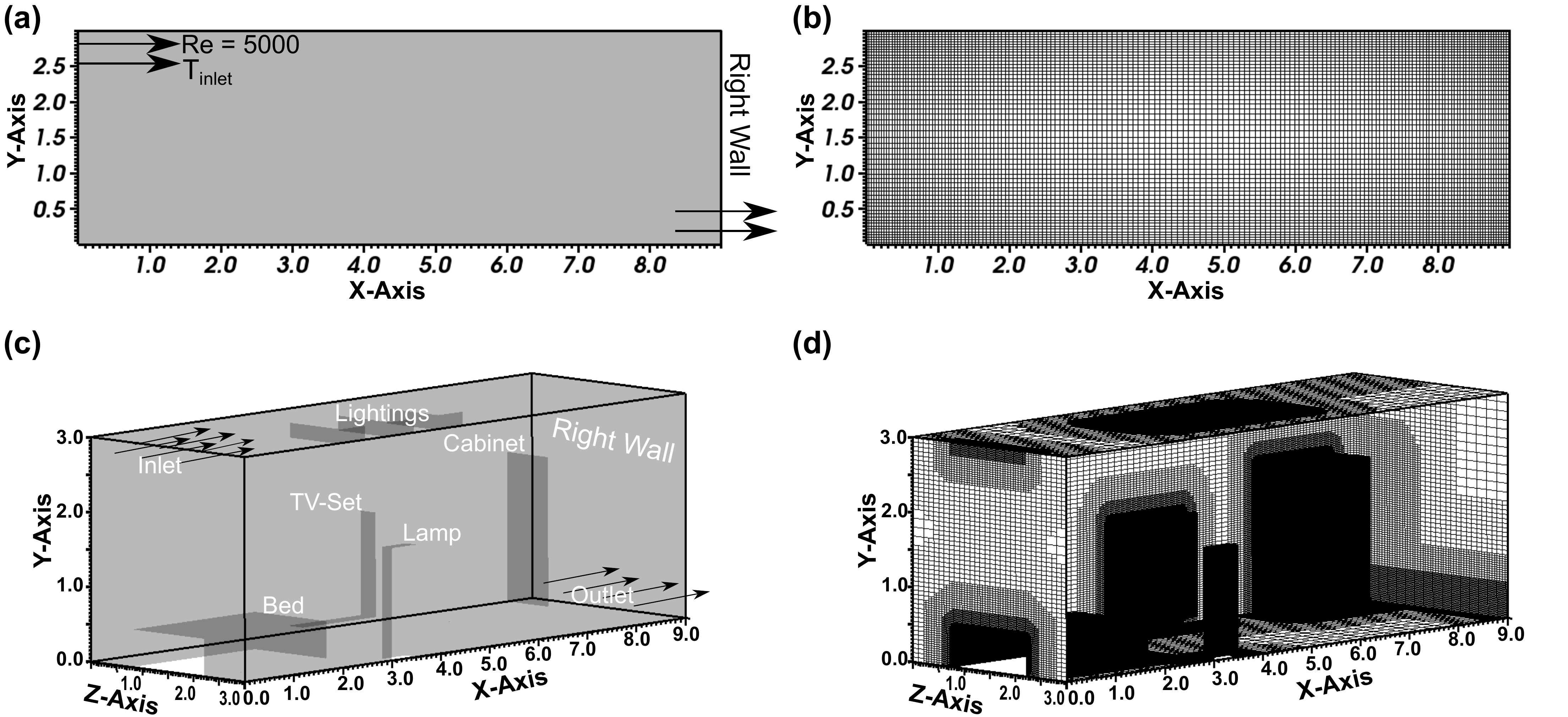}
\caption{Example problems (a) IEA-Annex 20 2D benchmark problem (b) 2D-Computational mesh used for the CFD simulations and PF operator construction (c) 3D-furnished geometry with different boundaries (d) 3D-Computational mesh used for the CFD simulation and PF operator constructions.}
\label{Fig:2d3dGeo}
\end{figure}
We consider the IEA-Annex20 \citep{Nielsena} benchmark problem in 2D and a furnished room in 3D. Fig~\ref{Fig:2d3dGeo} shows the 2D geometry and the 3D complex building and the associated computational mesh. These meshes were chosen after performance of rigorous mesh convergence analysis. 2D and 3D simulations were performed on meshes with 9462 and 0.6M cells, respectively. 

\begin{figure}[h]
    \centering
    \includegraphics[scale=0.28]{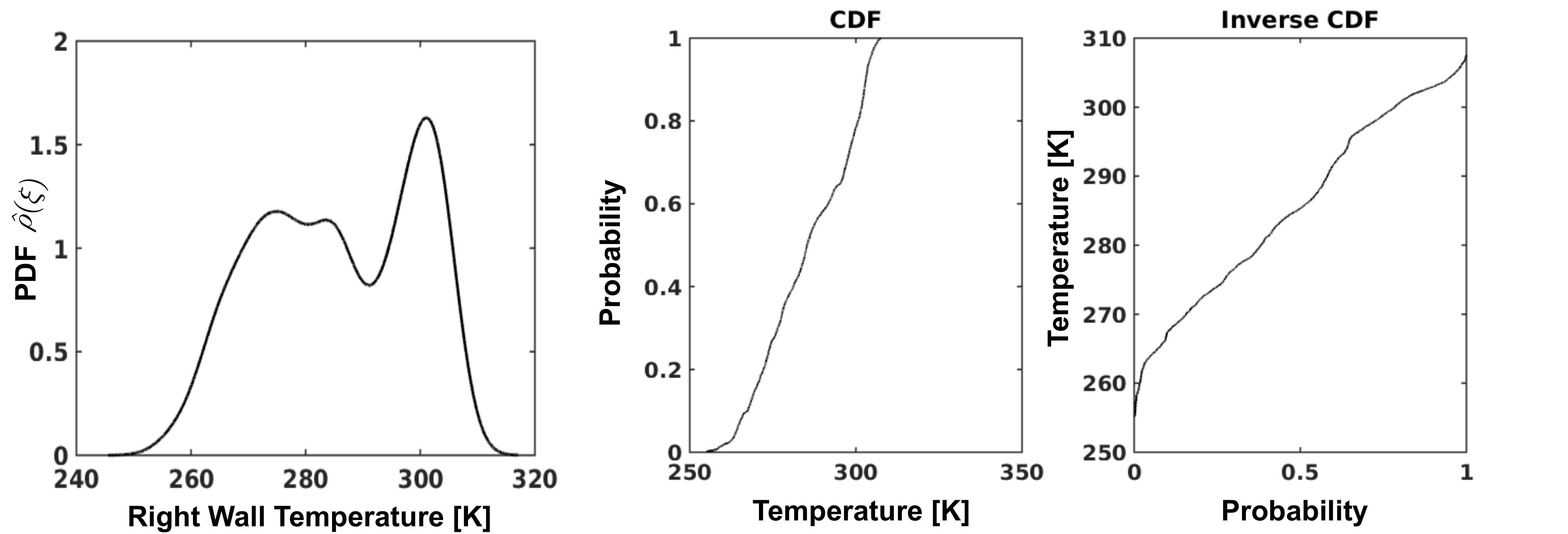}
    \caption{Approximated Probability density function, associated cumulative distribution function and inverse CDF for the right wall temperature of the building.}
    \label{fig:PDF_CDFRightWall}
\end{figure}

We consider a scenario where the boundary condition (i.e. wall temperature) for the right wall (for both 2D and 3D) is uncertain. The right wall temperature is considered to be a random variable, $\pmb{\xi}$. We construct the distribution using a data-driven, location-specific approach as detailed in Sharma et. al~\cite{sharma2018surrogate}. We use the TMY weather data for Des Moines, IA, U.S.A and perform a whole year simulation for an identical domain in the building simulation software EnergyPlus~\cite{Crawley2001a}. We extract the right wall temperature from the output logs of the simulation. These are realizations of the random variable, $\pmb{\xi}$. We then utilize standard techniques -- kernel density estimation \cite{bowman1997applied} -- to fit a probability distribution to the data. Fig.~\ref{fig:PDF_CDFRightWall} shows the data-driven probability distribution of the right wall temperature, as well as the associated cumulative distribution function and the inverse cumulative distribution function.

\subsection{Selecting the samples and checking for convergence in number of samples}\label{subsec:ParaSpace}
The next step is to construct the flow realizations by selecting samples of the random variable, $\pmb{\xi}$ (right wall temperature). A finite number of samples (and their associated probability) of the random variable are chosen so that they faithfully represent the true distribution of the random variable. The sampling of $\pmb{\xi}$ is carried according to the procedure outlined in appendix-\ref{appen}. We choose the number of samples to be $M = 7$. Since our final metric is the expected coverage, we perform a rigorous convergence analysis by increasing the number of samples, $M$, and evaluate the expected coverage for each $M$, until there is insignificant change in result ($\leq 4\%$). The results of this convergence analysis is shown in Table~\ref{Table:samplingConverg}, and indicates that $M = 7$ samples produces a sufficiently accurate representation of the random variable. While this convergence analysis was performed for the 2D case, our {\it a posteriori} analysis of the 3D results also indicated that this was a sufficient number of samples for the 3D case.
\begin{table}[h!]
\centering
\begin{tabular}{|c|c|c|}
\hline 
No. of Samples & Sample value on CDF & Error = $\frac{\parallel \mathbb{E}[\mathbb{V}_{i}]- \mathbb{E}[\mathbb{V}_{5}]\parallel}{\parallel \mathbb{E}[\mathbb{V}_{5}] \parallel}$ \\ \hline 
2 & 0.0, 1.0 & 0.4434 \\ 
\hline 
3 & 0,0.5,1.0 & 0.2418 \\ 
\hline 
5 & 0,0.3,0.5,0.7,1.0 & 0.1172 \\ 
\hline 
7 & 0,0.1,0.3,0.5,0.7,0.9,1.0 & 0.0368 \\ 
\hline 
9 & 0,0.1,0.3,0.4,0.5,0.6,0.7,0.9,1.0 & - \\ 
\hline 
\end{tabular} 
\caption{Convergence analysis of the expected coverage $\mathbb{E}[\mathbb{V}]$ computed for increasing number of samples of the random variable.}
\label{Table:samplingConverg}
\end{table}

\subsection{Simulation of the flow fields and construction of the PF Operators} \label{subsec:FlowfieldPF}
We perform non-isothermal simulations in 2D and 3D for each of the $M = 7$ cases identified. We utilize the OpenFOAM \citep{Jasak2007} simulation tool, specifically the \textit{buoyantboussinesqsimpleFoam} solver. We set a convergence threshold of $10^{-5}$ to solve the coupled Navier Strokes and energy equations. All walls have no-slip boundary conditions, with velocity inlet and zero-gradient outlet conditions for both 2D and 3D problem. All the walls except the right wall are considered well insulated. The inlet air temperature and the right wall temperature for each of the $M = 7$ sampling points are listed in Table-\ref{Table:SamplingBC}. The inlet air temperature is chosen such that indoor thermal comfort is maintained \cite{ridouane2011evaluation}. 

\begin{figure}[hp]
\centering
\includegraphics[scale=0.35,width=\textwidth,height=0.95\textheight,keepaspectratio]{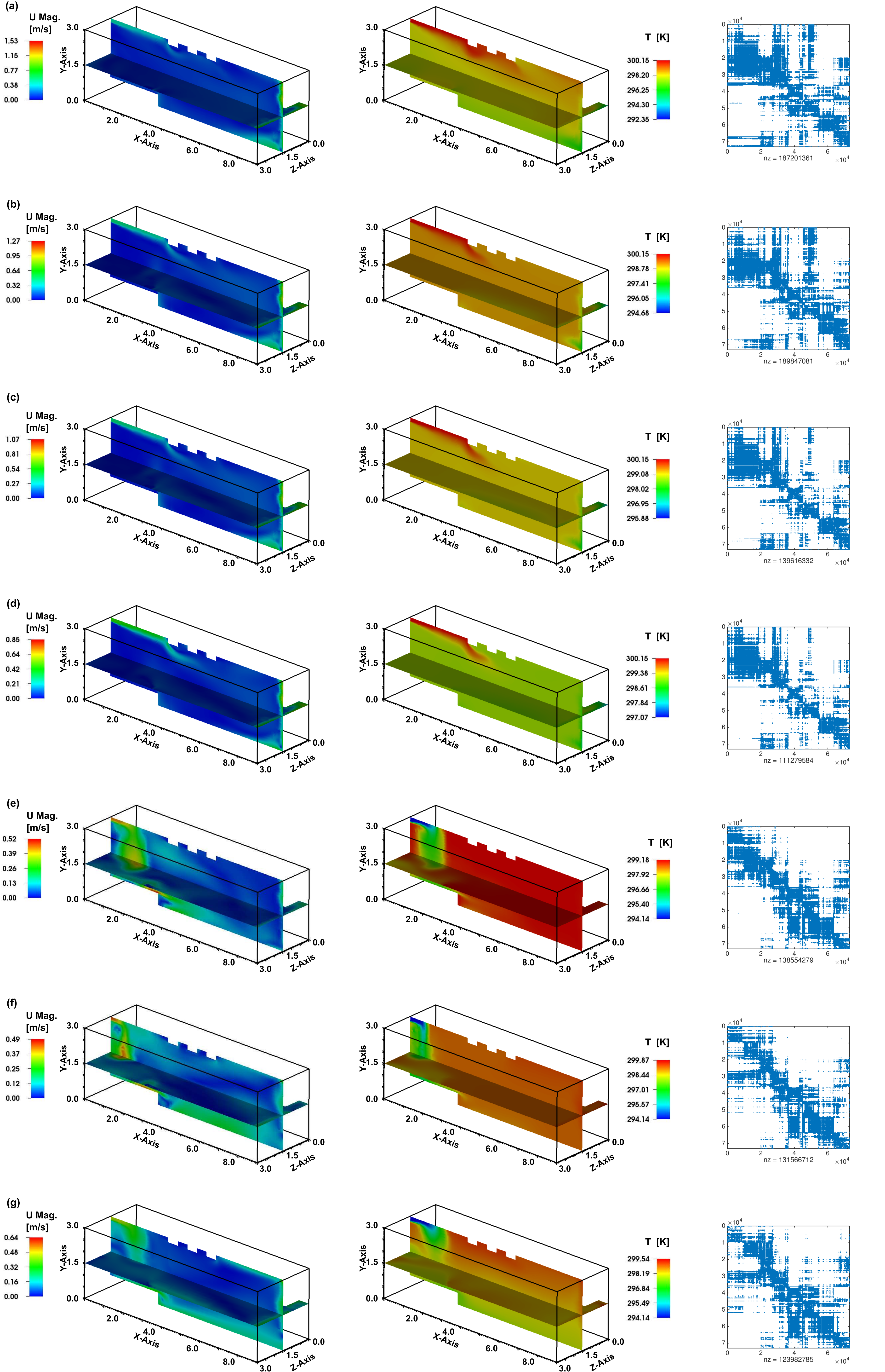}
\caption{Velocity magnitude contours for boundary conditions sampled for seven points on CDF for a 3D geometry (a)CDF-0.0 (b)CDF-0.1 (c)CDF-0.3 (d)CDF-0.5 (e)CDF-0.7 (f)CDF-0.9 (g)CDF-1.0}
\label{fig:3DFlowFieldPF}
\end{figure}

\begin{table}[h!]
\centering
\begin{tabularx}{.8\textwidth}{|X|X|X|}
\hline 
Sample points on Right wall Temperature CDF & Right wall Temperature [K] & Inlet Temperature $T_{inlet}$ [K] \\ 
\hline 
0.0 & 255.0 & 300.0 \\ 
\hline 
0.1 & 267.3 & 300.0 \\ 
\hline 
0.3 & 276.4 & 300.0 \\ 
\hline 
0.5 & 285.3 & 300.0 \\ 
\hline 
0.7 & 297.2 & 294.0\\ 
\hline 
0.9& 302.9 & 294.0\\ 
\hline 
1.0 & 307.5 & 294.0\\ 
\hline 
\end{tabularx}
\caption{The Boundary conditions associated with the sampling points to compute the steady flow field and construct the PF operator.}
\label{Table:SamplingBC}
\end{table}
The computational mesh used for the numerical simulation significantly affects the results and hence grid convergence study is carried for both 2D and 3D geometries (please see appendix.~\ref{sec:GridConv} for detailed mesh convergence results). The study resulted into the choice of 9462 cells in 2D and 0.6 Million cells for the 3D geometry. Following the computation of the velocity fields, we next turn to the construction of the PF matrix. As discussed in section-\ref{subsec:ConsPFOpertor} the size of the PF matrix is depend on the number of states/cells used in computation of flow field. We choose not to use the fine scale mesh used for CFD simulations to construct the PF matrix for each realization\footnote{Using the CFD mesh would result in a PF matrix of size $0.6\text{M}\times 0.6\text{M}$. Such a large matrix will require a lot of memory to store, and subsequent analysis will be infeasible}. We project the computed flow fields from the CFD mesh onto a coarser mesh containing around 70K cells. Therefore, the PF operator constructed is of size $70\text{K}\times70\text{K}$. A solver \textit{makeMarkovMatrixSteadyDiff} is developed in OpenFOAM to construct the PF matrix by using the flow field obtained after solving NS equations. We make this solver freely available at \url{git@bitbucket.org:baskargroup/markovmatrixgeneration.git}.
\\
\begin{figure}[h]
\centering
\includegraphics[scale=0.5,width=\textwidth]{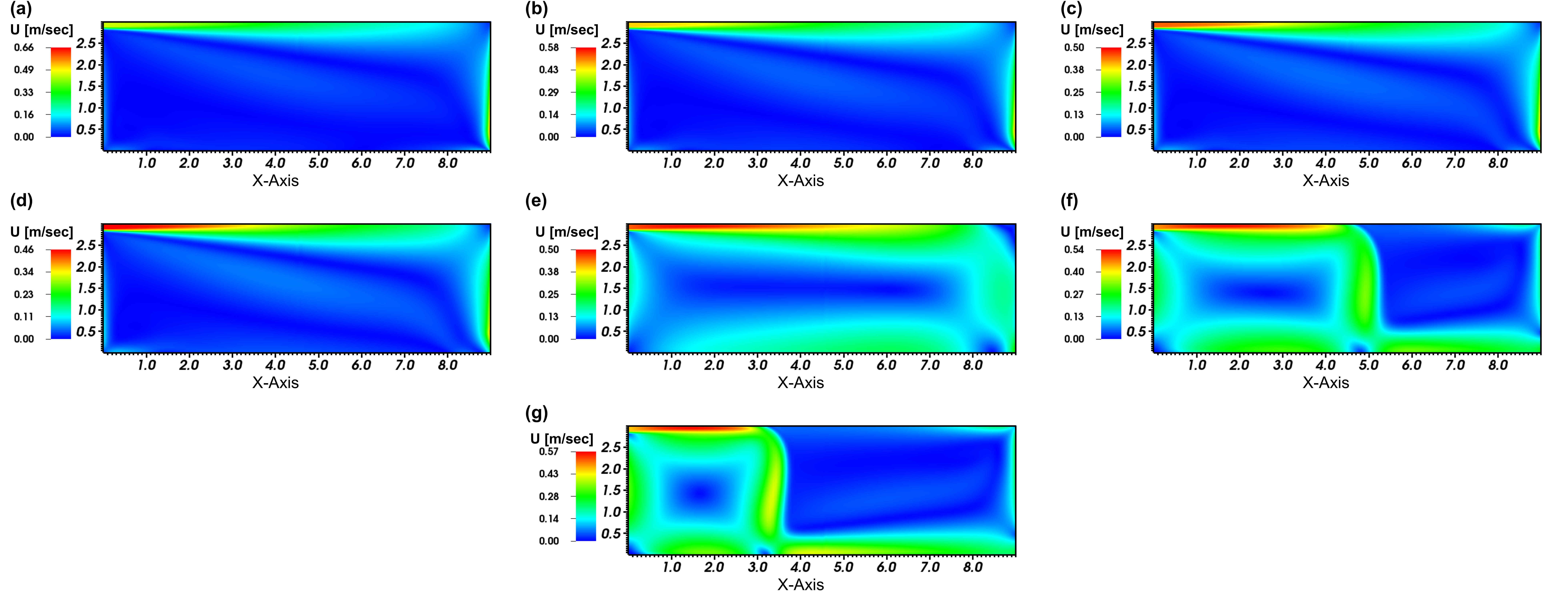}
\caption{Velocity magnitude contours for boundary conditions sampled for seven points on CDF for a 2D geometry (a)CDF-0.0 (b)CDF-0.1 (c)CDF-0.3 (d)CDF-0.5 (e)CDF-0.7 (f)CDF-0.9 (g)CDF-1.0}
\label{fig:2DFlowFieldPF}
\end{figure}
\indent Figure-\ref{fig:2DFlowFieldPF} shows the 2D velocity field contours computed by solving the steady NS equations for the sampled seven boundary conditions. It can  seen that various flow conditions are observed in the building by simply changing the right wall boundary conditions. The PF operator for each of the $M = 7$ cases is used in the sensor placement for the 2D problem. Figure-\ref{fig:3DFlowFieldPF} shows contours of flow field and temperature on the mid-plane cross-sections along with the sparse matrix representation of the PF operator. These PF operators are used to find the sensor locations for the 3D geometry on the expectations based sensor placement framework developed in section-\ref{sec:MethodS}.

\subsection{Sensor placement accounting for uncertainty: 2D case}\label{subsec:2D case}
\subsubsection{Validation of the contaminant transport via PF operator
2D}\label{subsubsec:Validation2D}
\begin{figure}[h]
\includegraphics[scale= 0.5,width=\textwidth]{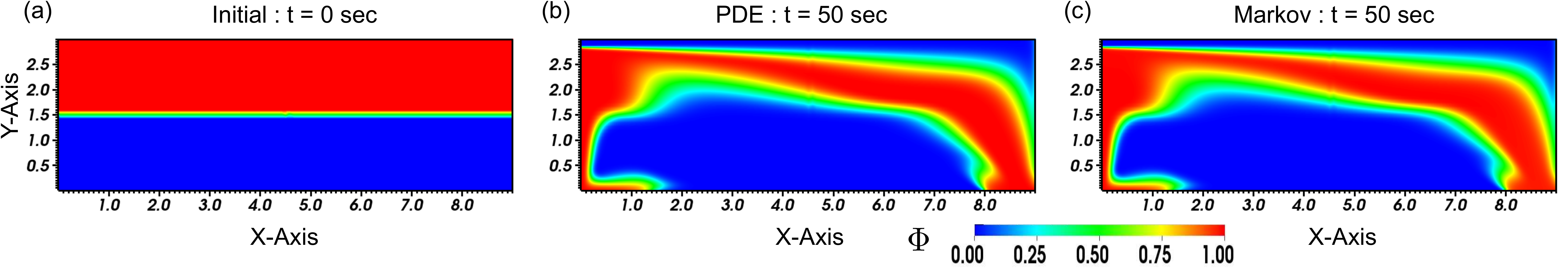}
\caption{Contaminant Transport comparison of PDE and Markov Matrix in 2D (a) The initial contaminant concentrations at t=0 used for initializing both PDE and Markov. (b) The iso-surface of the contaminant transport by PDE after t=50 sec (c) The contaminant tranport iso-surface by PF operator after t=50sec.}
\label{fig:ConcenCompare}
\end{figure}

\begin{figure}[h]
\centering
\includegraphics[scale=0.5,width=0.75\linewidth]{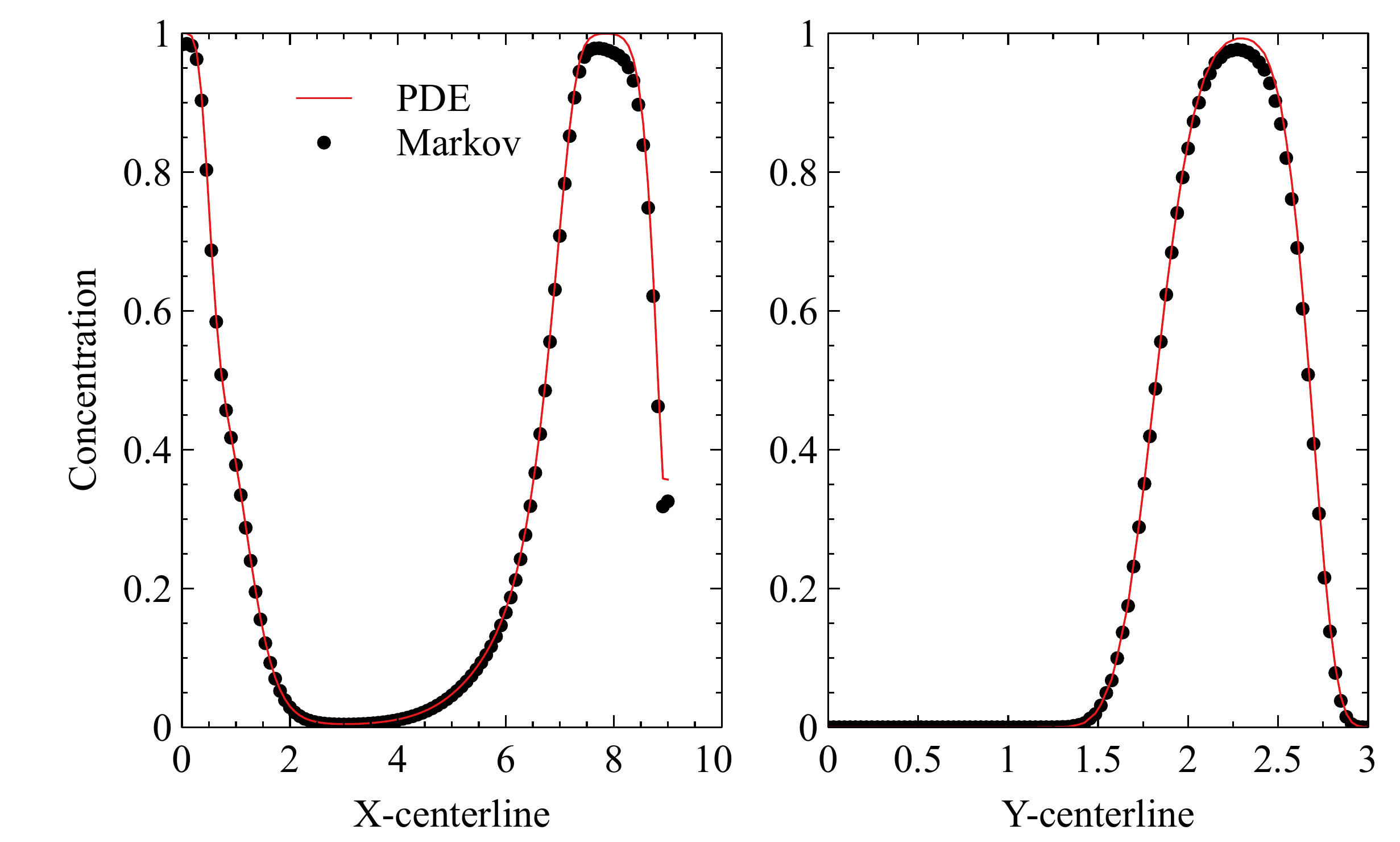}
\caption{Contaminant concentration profile comparison on the x-centerline and y-centerline of PDE and Markov approach}
\label{fig:2DConceprofile}
\end{figure}

To validate the constructed PF operators we compare the contaminant transport, $\Phi$, computed by the PDE Eq.~(\ref{Eq:scalarTran}) and Markov matrix Eq.~(\ref{Eq:3}). While we validate each of the $M = 7$ PF matrices against a PDE based result, we show results for a single case for brevity. Figure-\ref{fig:ConcenCompare}(a) shows the initialized scalar map where red represents a scalar concentration of one and blue corresponds to zero. The scalar transport PDE equation is solved to a time horizon of 50 sec using the OpenFoam solver, and compared with the Markov approach. Figure~\ref{fig:ConcenCompare}(a-b) compares the scalar concentration contours after 50 sec by PDE and Markov approach. It can be seen that they closely match each other. The concentration profiles are also plotted  and compared  about the mid-planes of the building which is shown in fig.\ref{fig:2DConceprofile}. The Markov results are represented as dots, and very closely match the PDE based predictions with an $L_2$ error of $10^{-4}$. The validation shows the accuracy of the Markov approach and effectiveness of the matrix-vector product based approach in predicting the transport of scalar concentration.

\subsubsection{Sensor locations with no-constraints}
\begin{figure}[ht]
\centering
\includegraphics[scale=0.5,width=0.95\linewidth]{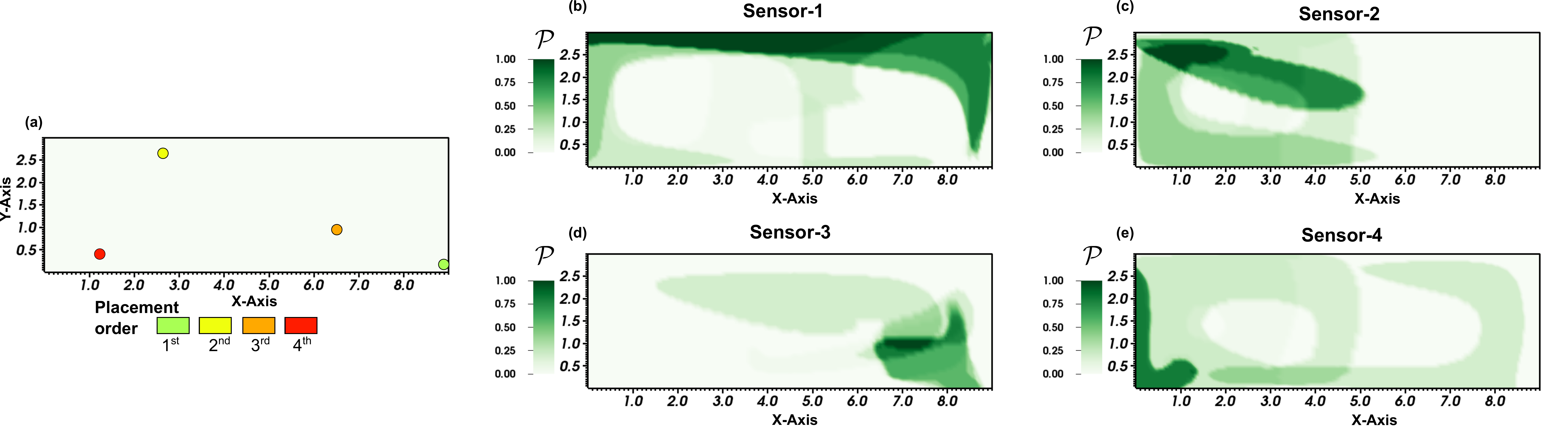}
\caption{(a) The optimal sensor locations accounting for uncertainty for the 2D building (b-e) The probabilistic $\mathcal{P}$ coverage map for 1$^{st}$,2$^{nd}$,3$^{rd}$,4$^{th}$ sensor.}
\label{fig:2DSensorLoc}
\end{figure}

\begin{figure}[h]
    \centering
    \includegraphics[scale=0.5, width = 0.75\textwidth]{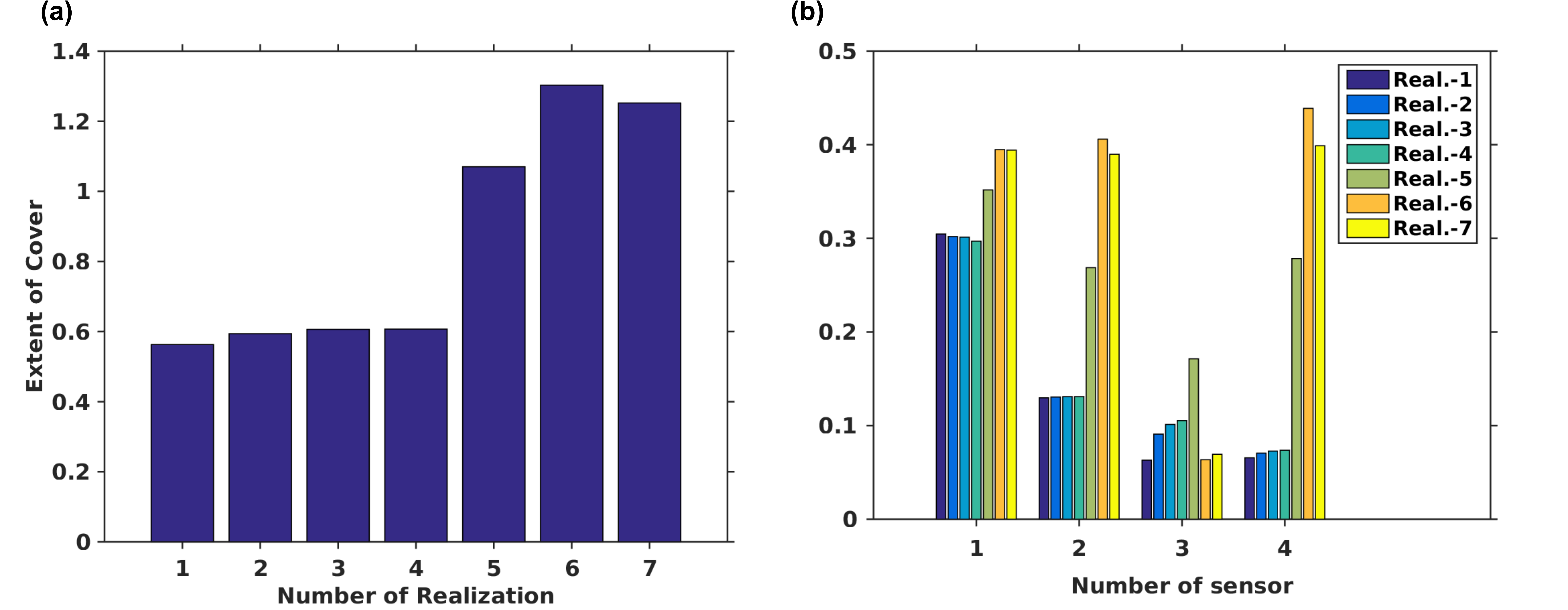}
    \caption{(a-b) The extent of cover for the four sensors placed and volume coverage of each sensor associated with each sample point.}
    \label{Fig:2DSensorCoverage}
\end{figure}

We identify sensor locations under uncertain conditions for the no constraint scenario. We place four sensors in the domain with a sensing time of $t= 60 sec$. The accuracy of the sensor used for thresholding the contaminant matrix is $\epsilon_{acc} = 0.01\%$. The optimal sensor location are shown in Fig.~\ref{fig:2DSensorLoc}(a). The probability coverage for each sensor is shown in Fig.\ref{fig:2DSensorLoc}(b-e). 
For the sensors the associated volumetric coverage for the sampled seven realization and the extent of cover is shown in Fig. \ref{Fig:2DSensorCoverage}(a-b). The total expected coverage for all the realization using four sensors is  83.65\%. Having illustrated the approach for a 2D benchmark, we next deploy this approach for a complex 3D scenario. 


\subsection{Sensor placement accounting for uncertainty for a complex building: 3D case}\label{subsec:3D case}
Next we discuss the results of sensor placements for the 3D complex furnished building geometry. The constructed PF operators and the flow fields are already shown in section-\ref{subsec:FlowfieldPF}. We first validate the accuracy of the PF operator by comparing PF based contaminant transport predictions with PDE based predictions.

\subsubsection{Validation of PF based contaminant transport}
\begin{figure}[h]
\includegraphics[width=\textwidth,height=\textheight,keepaspectratio]
{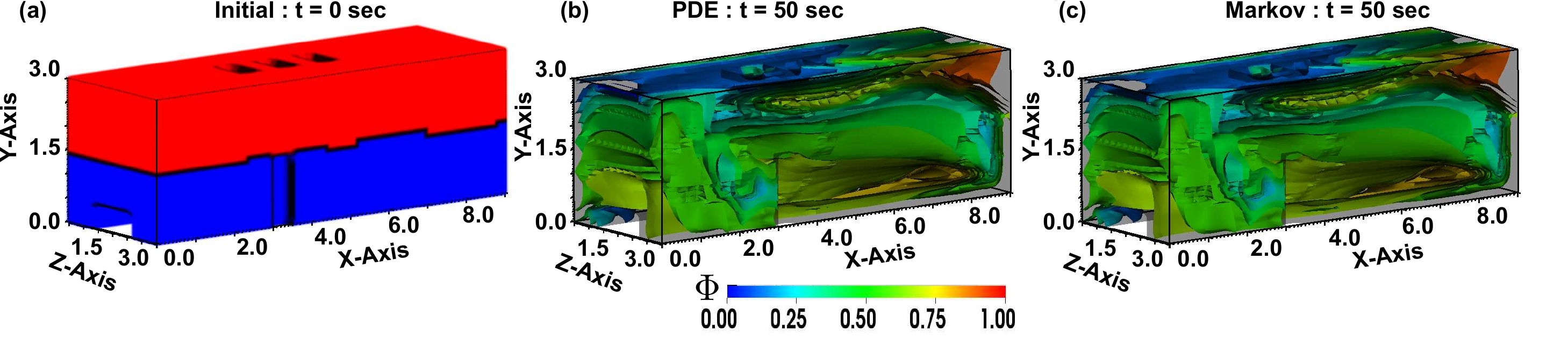}
\caption{Contaminant Transport comparison of PDE and Markov Matrix in 3D (a) The initial contaminant concentrations at t=0 used for initializing both PDE and Markov. (b) The iso-surface of the contaminant transport by PDE after t=50 sec (c) The contaminant tranport iso-surface by PF operator after t=50sec.}
\label{fig:3DPFContaminantvalidations}
\end{figure}

\begin{figure}[h]
\centering
\includegraphics[width=0.75\linewidth]
{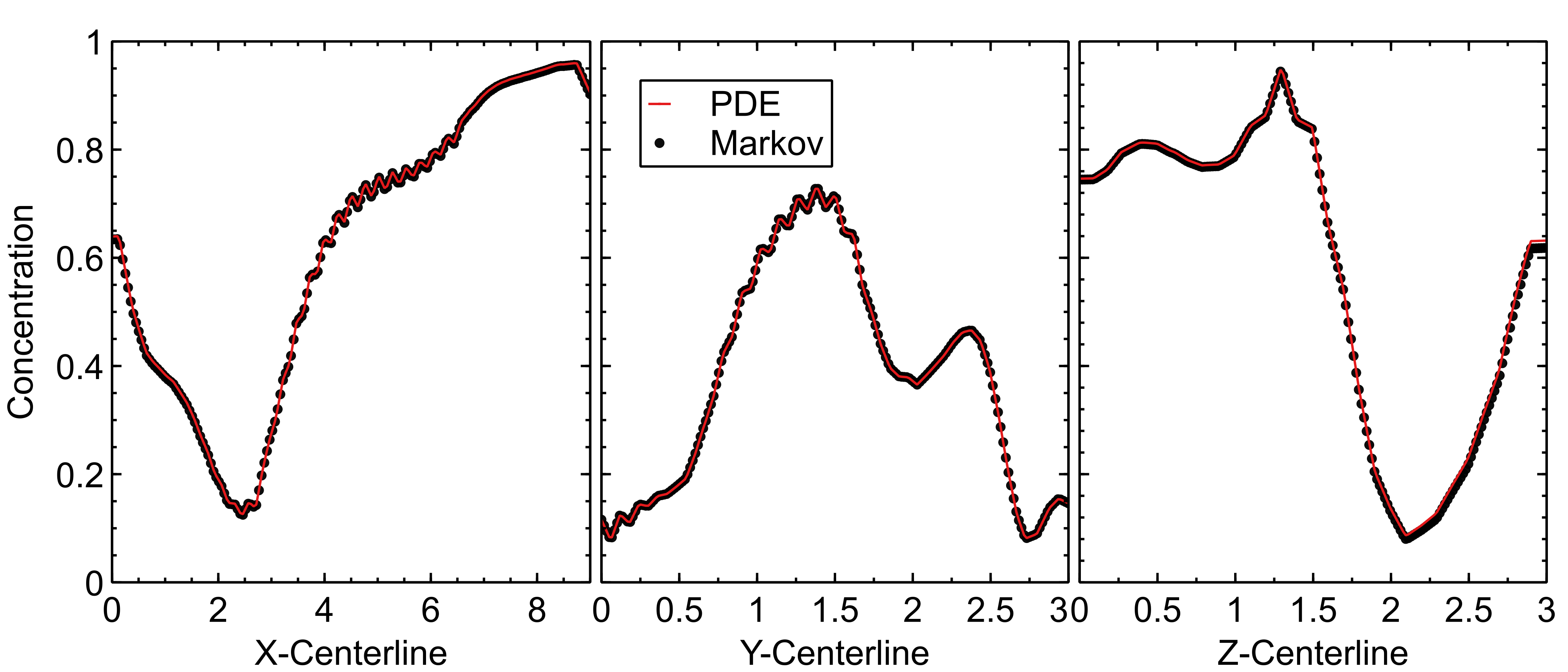}
\caption{The contaminant concentration along the centerlines passing through the three mid planes of the geometry is compared for the PDE prediction and the Markov prediction.}
\label{fig:3PF_PDEprofilecomp}
\end{figure}

We perform validation similar to the 2D PF operator validation case discussed in section-\ref{subsubsec:Validation2D}. The transport of contaminant, $\Phi$, is performed using both PDE solutions and PF operator. The results are shown for one of the constructed PF operator but all the constructed PF operators are validated in similar manner before using them for sensor placement. Figure-\ref{fig:3DPFContaminantvalidations}(a) shows the initial distribution of the contaminant at t=0 where the contaminant density of one is marked with red and zero where the contaminant is absent. Figure-\ref{fig:3DPFContaminantvalidations}(b) shows iso-surface of the evolution of the contaminant from the solution of PDE in the domain at t=50sec. Figure-\ref{fig:3DPFContaminantvalidations}(c) show the iso-contours of the contaminant by the PF operator at t=50 sec. It can be observed that two iso-contours are indistinguishable. To further validate the PF operator, we monitor the concentration profile along the three center axes passing, as shown in Figure-\ref{fig:3PF_PDEprofilecomp}. It can be seen that they exactly overlap each other which shows the accuracy of the PF operator in transporting the contaminant. 

\subsubsection{Sensor placement with no constraints}
\begin{figure}[h]
\includegraphics[width=\textwidth]
{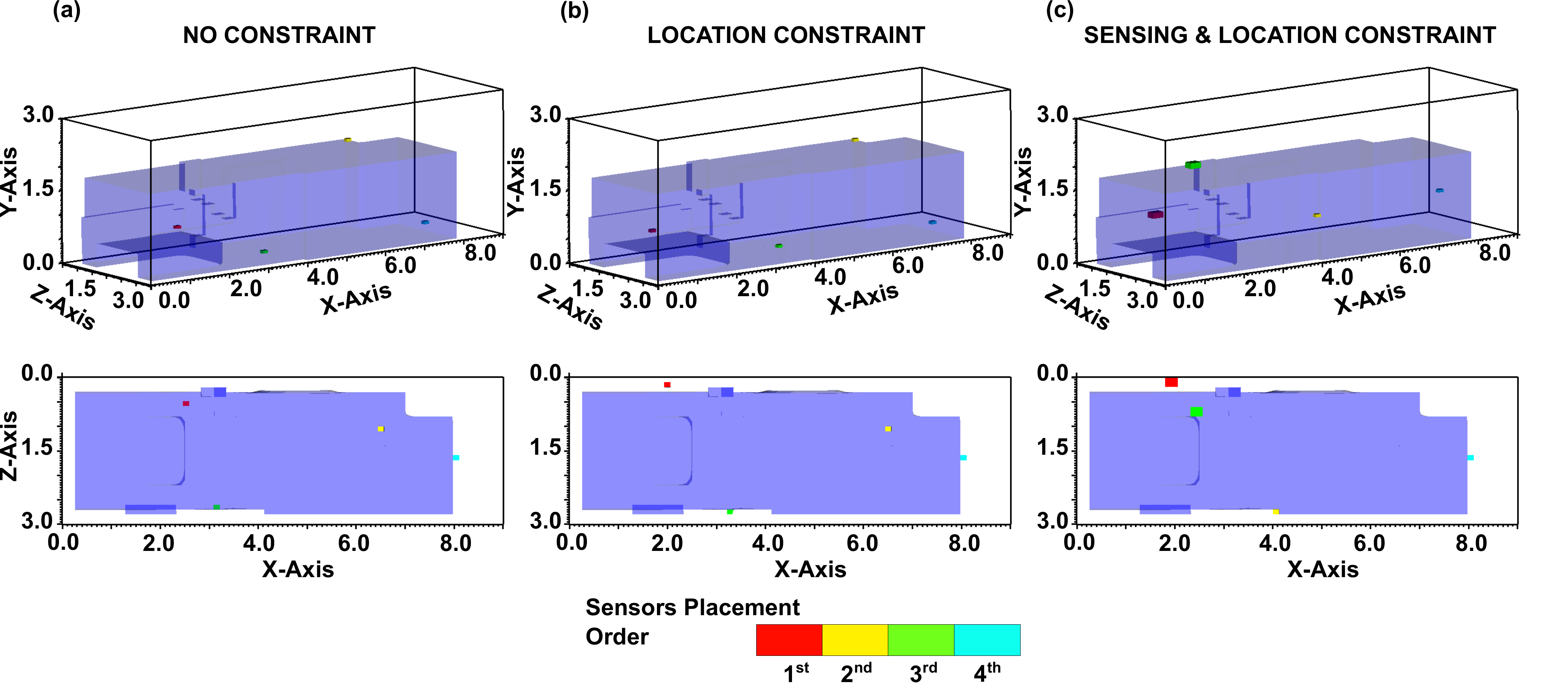}
\caption{The sensor locations are shown for three scenarios with isometric and top view. The blue region is the occupied zone for the building. (a) In the no constraints case the sensor are free to be placed anywhere inside the domain. (b) In the case of location constraints, the sensors are placed only outside the occupied zone (c) In the case of sensing and location constraints, the sensors are placed outside the occupied zone but must sense the occupied space.}
\label{fig:3DSensorloc}
\end{figure}

We first place a fixed number ($k = 4$) of sensors to maximize the expected coverage without any sensing and location constraints. Fig.\ref{fig:3DSensorloc}(a) shows the sensor location under this case. It can be seen that to maximize the expected coverage the first and third sensors are placed inside the occupied region. Another interesting observation is related to the placement of the first sensor. For the deterministic case, the first sensor is usually placed close to the outlet, as has been extensively reported by \citep{Fontanini2016}. Placing the first sensor close to the outlet maximizes the observability of the domain, as all the air (and contaminant) must exit from the outlet. In contrast to the deterministic case, maximization of the expected coverage is not achieved by placing the first sensor near the outlet but (for this geometry) is inside the occupied zone of the building. 

\subsubsection{Sensor placement under location constraints}
The location constraints are particularly attractive for critical applications such as detection of CBW (chemical and biological weapons) and TID (transmission of infectious diseases). As most of the CBW and TID sensors are expensive and can be damaged if placed in the occupied space, this constraint forces the sensors to be placed outside the occupied zone. Fig.\ref{fig:3DSensorloc}(b) shows the obtained sensor location for four sensors. 

\subsubsection{Sensor placement with sensing and location constraints}
In case of indoor air quality applications, the sensing of the contaminant concentrations
in the occupied zone is much important than outside the occupied zone. Therefore a sensing constraint is added while determining the optimal location of the sensor. Similar to the case of CBW and TID scenarios, it is important that the sensor must not be placed where the occupant can affect or damage it. For this reason a location constraint is also incorporated so that the sensor must be placed outside the occupied zone. Fig.\ref{fig:3DSensorloc}(c) shows the sensor location in the isometric view and top view. In this case sensors tend towards to be placed near the occupied zone or at the outlet. 

\subsubsection{Discussion of quantitative measures}
 \indent We finally compute quantitative measures for the placed sensors. Under the no-constraint scenario, the optimal placement of sensors results in an expected coverage of 31.4 \% of the total volume. That is, under all possible flow condition scenarios produced by an uncertain wall boundary condition, the sensors are expected to cover 31.43 \% of the domain. Interestingly, under location constraints, the expected coverage does not change much from the no-constraint case. This is primarily because of the increase in coverage sensor-2 provides for realization-2. With the addition of both location and sensing constraints there is a  decrease in the expected coverage to 15.0 \% of the total volume. However, since in this case we are only interested in the occupied zone, the sensor placed with sensing and location constraints cover 21.50 \% of the {\it occupied space}. 

\begin{figure}[p]
\centering
\includegraphics[scale=0.22]{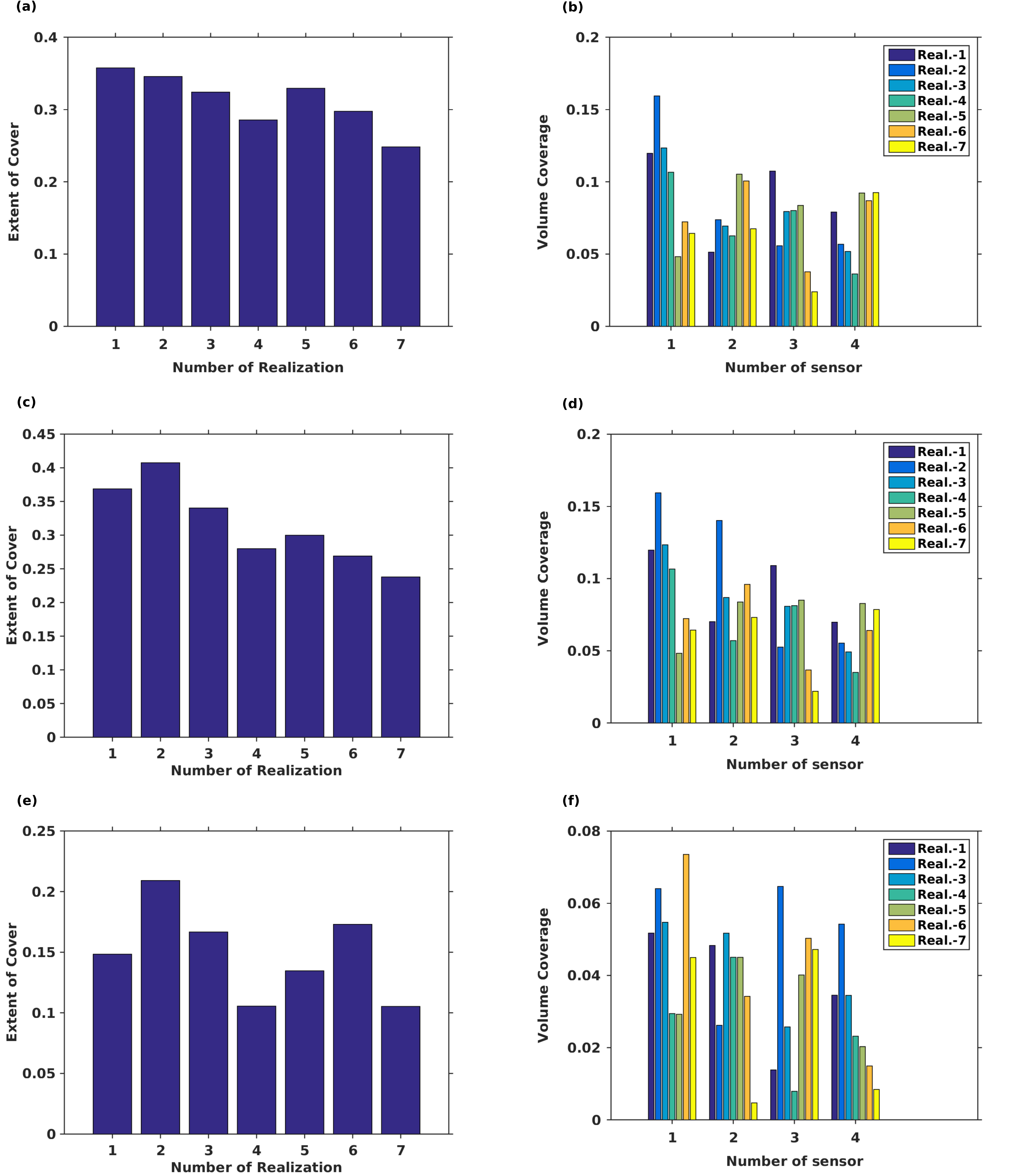}
\caption{(a-b)The extent of cover for the four sensors placed without any placement constraint and volume coverage of each sensor associated with each sample point in the study. (c-d) The extent of cover for the four sensors placed with location constraints for placement and volume coverage of each sensor associated with each sample point in the study. (e-f) The extent of cover for the four sensors placed  with a constraint on location and sensing and  volume coverage of each sensor associated for each sample point in the study.}
\label{fig:CoverageMap_ExtentCoverBar}
\end{figure}

For each of the $M = 7$ realizations, we compute the volumetric coverage for each realization associated with each sensor. Figure-\ref{fig:CoverageMap_ExtentCoverBar} shows plots of these quantitative measures, where fig-\ref{fig:CoverageMap_ExtentCoverBar}(a-b) corresponds to the sensor placement with no-constraint on placement. (c-d) corresponds to the constraints on the location as well as sensing and (e-f) corresponds to the sensor placement with a only constraint on the location of the sensor. By comparing the extent of cover for fig-\ref{fig:CoverageMap_ExtentCoverBar}(a-c-d) it can be seen that the sensor positions obtained under no constraint result in high extent of cover for all realizations. While for the cases with constraint on placement the extent of cover varies significantly with each realization. Comparing the volume cover for fig-\ref{fig:CoverageMap_ExtentCoverBar}(e-d-f) it can be observed that under no constraint sensor placement the first sensor performs majority of sensing for each realization, while in case of sensing and location constraint the sensing is well distributed among the four sensors.   

\section{Conclusion}\label{sec:Conclude}
The work introduces a method for accounting for weather uncertainty during optimal sensor placement. The method is based on constructing the Perron-Frobbenious operator from the flow field using a set based approach. We use the theoretical background of uncertainty quantification to formulate the sensor placement problem. The weather affecting the flow field is accounted as uncertainty in the current work. We utilize a data-driven approach to construct the uncertain boundary condition as a probability distribution. We then select a finite number of representative samples from this distribution. The samples and their associated probability of occurrence are chosen such that the expected value of these samples is identical to the full distribution. Full CFD simulations are performed corresponding to this finite set of boundary conditions. The velocity fields are then used to construct individual PF operator which is used in the algorithm to identify the sensor layout for the domain.\\
\indent This approach can naturally account for any uncertainty to determine the optimal sensor layout. By using a dynamical systems approach we gain two key advantages: (a) we circumvent the numerical complexity and computational challenges with complementary optimization/inverse based approached. Additionally, incorporating effects of uncertainties is non-trivial in other methods, (b) the PF based approach produces a mathematically rigorous and {\it globally optimal} sensor layout. This is in contrast to other optimization based approaches which can only identify local optima. We illustrate the approach for problems in both 2D and 3D building geometry. We envision that the developed framework can be used for critical applications including sensor layout for CBW scenarios, transmission of infectious diseases, and indoor air quality. The contributions from this paper also provide ability for designers, engineers, and researchers to better understand the probabilistic coverage maps for each sensor location. 

\section*{Acknowledgement}
UV acknowledges partial support from NSF CAREER 1150405. BG acknowledges partial support from NSF CAREER 1149365.

\begin{appendices}  
\section{Appendix}
\subsection*{Sampling points and associated weights}\label{appen}
We use sampling theory to choose a finite set of realizations of a random variable, $\pmb{\xi}$, with a probability distribution, $\rho(\pmb{\xi})$.  Given an output variable, $\mathbf{v}(\pmb{\xi})$, that depends on the random variable, the expected value of this output variable is given by
\begin{equation}
\mathbb{E}[\mathbf{v}] = \int \mathbf{v}(\pmb{\xi}) \rho(\mathbf{\pmb{\xi}}) d\pmb{\xi}   
\label{Eq:expecalcu}
\end{equation}
Now, the finite number of realizations, $\{\pmb{\xi}_1, \cdots, \pmb{\xi}_M \}$ are chosen such that they can accurately recreate the expected value. That is
\begin{equation}
\int \mathbf{v}(\pmb{\xi}) \rho(\mathbf{\pmb{\xi}}) d\pmb{\xi} = \sum_{i=1}^{M}\mathbf{v}(\pmb{\xi}_i)\theta_i
\label{Eq:expecalcu}
\end{equation}
where $\mathbf{v(\pmb{\xi})_i}$ are the outputs evaluated at the sampling points $\pmb{\xi}_i$, and $\theta_i$ are the probabilistic weights (defined below). We use the approach detailed in Zabaras and Ganapathysubramanian~\cite{zabaras2008scalable} to sample from a distribution\footnote{In a data-driven context, we do not necessarily have a parametrized distribution for $\pmb{\xi}$. The usual approach is to use a kernel density estimator to construct a non-parametric representation of the distribution from data.}. Essentially, we sample points uniformly from the inverse cumulative distribution function (iCDF) of $\pmb{\xi}$ and increase the number of points until the expected value has converged. This ensures that we have chosen representative points which provides a statistical representation of the actual distribution of the random variable. In case of more number of random variable, sampling techniques such as latin-hypercube sampling (LHS),importance sampling and sparse grid collocation can be used~\cite{ganapathysubramanian2007sparse}.       

The determination of the weights $\theta_i$ is performed using interpolation theory. The output variable, $\mathbf{v}$, which is a function of $\pmb{\xi}$ is represented as a sum of basis functions $N_i(\pmb{\xi})$ (defined over the support of $\pmb{\xi}$). That is
\begin{equation}
\mathbf{v}(\pmb{\xi}) \approx \sum^{M}_{i=1} \mathbf{N}_{i}(\pmb{\xi})\mathbf{v}_{i}
\label{Eq:CoverageApprox}
\end{equation}
where $\mathbf{v}_i$ are the outputs evaluated at the points $\pmb{\xi}_i$. 
Inserting this representation into Eq.~\ref{Eq:expecalcu} results in
\begin{equation}
\begin{split}
\mathbb{E}[\mathbb{V}] &= \int  \sum^M_{i=1} \mathbf{N}_{i}(\pmb{\xi})\mathbf{v}_{i} \rho(\mathbf{\pmb{\xi}}) d\pmb{\xi}  \\
&=  \sum^{M}_{i=1} \mathbf{v}_{i} \int  \mathbf{N}_{i}(\pmb{\xi}) \rho(\mathbf{\pmb{\xi}}) d\pmb{\xi}\\
&= \sum^{M}_{i=1} \mathbf{v}_{i} \theta_{i}
\end{split}
\label{Eq:Subsexpecalcu}
\end{equation}
Thus, $\theta_i = \int  \mathbf{N}_{i}(\pmb{\xi}) \rho(\mathbf{\pmb{\xi}}) d\pmb{\xi}$. \\
\indent To illustrate the approach we consider $\pmb{\xi}$ to be a standard normal variable exhibiting a Gaussian distribution $(\mathcal{N}(0,1),\mu = 0.5,\sigma=0.05)$. We evaluate $\theta_i$ and compare the true integral (Eq.~\ref{Eq:CoverageApprox}) with the approximate evaluations using $M = 7$ sampling points, for three different functional forms of $\mathbf{v}$. Linear basis functions were used to approximate $\mathbf{v}(\pmb{\xi})$ in Eq.(\ref{Eq:Subsexpecalcu}). Table-\ref{Table: ExpectedCalculations} show the comparisons with three different functions. It can be seen that the approximation integral closely matches the exact integral up to two decimal places. 
\begin{table}[H]
\centering
\begin{tabular}{|c|c|c|}
\hline 
$\mathbf{v}(\pmb{\xi})$ & Exact Integral & Approximate Integral  \\ 
\hline 
$\mathbf{v}(\pmb{\xi}) = \pmb{\xi}$& 0.5
 & 0.499 \\ 
\hline 
$\mathbf{v}(\pmb{\xi}) = \pmb{\xi}^2$ & 0.252 & 0.259 \\ 
\hline 
$\mathbf{v}(\pmb{\xi}) = exp(\pmb{\xi})$ & 1.650 & 1.656 \\ 
\hline 
\end{tabular}
\caption{The comparison of the expectation integral $\mathbb{E}[\mathbb{V}]$ by the exact evaluation using Eq.(\ref{Eq:expecalcu}) and Eq.(\ref{Eq:Subsexpecalcu}) are presented in the table.}
\label{Table: ExpectedCalculations}
\end{table}

\section{Grid Convergence}\label{sec:GridConv}
\begin{figure}[H]
\centering
\includegraphics[scale=0.35,width=0.8\linewidth]{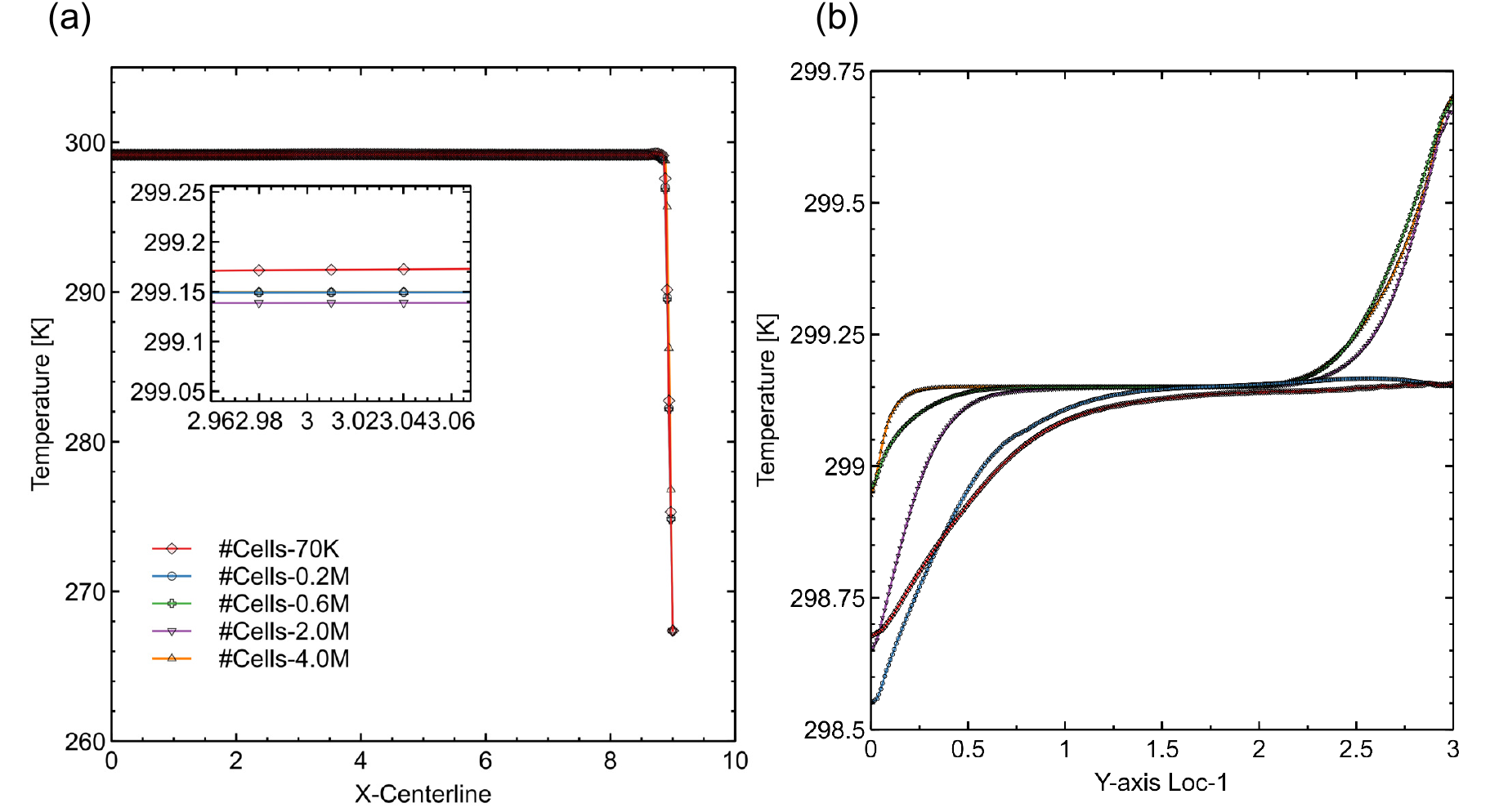}
\caption{Grid independence study for 3D numerical computation (a) Temperature profile comparison of arious grids on x-centerline (b) Temperature profile comparison of various grid along y-axis close to the cabinet in the 3D geometry.}
\label{fig:GridConv}
\end{figure}

 The grid convergence study is performed  using cdf-0.1 boundary conditions shown in table-\ref{Table:SamplingBC} for the 3D geometry. The temperature profiles are shown in Fig.\ref{fig:GridConv} plotted along the x-axis passing through the room center and along an arbitrary chosen y-axis close to cabinet ((7.5,0,0.5),(7.5,3.0,0.5)) for the 3D geometry(fig.~\ref{Fig:2d3dGeo}). It can be seen that for all the meshes the X-centerline profile closely overlap each other. While in case of Y-line the difference are merely $\pm$ 0.2 K between the fine 4.0M mesh and the chosen 0.6M mesh. Therefore, all the CFD  computations to compute the flow field were carried using 0.6M mesh.
\end{appendices}

\bibliographystyle{model1-num-names}
\bibliography{refs,references}

\mbox{}
\nomenclature{${U}$}{steady/unsteady flow field}
\nomenclature{$\mathbf{P}$}{Markov Matrix}
\nomenclature{$\Omega$}{computational domain}
\nomenclature{$\Phi$}{continuous form scalar contaminant}
\nomenclature{$S_{\Phi}$}{source of contaminant}
\nomenclature{$\hat{S}_{t_i,t_{i+1}}$}{ discrete form of source location}
\nomenclature{$t$}{time [sec]}
\nomenclature{$t_{f}$}{final time of simulation [sec]}
\nomenclature{$\delta t$}{time associated with Markov matrix}
\nomenclature{$\tau$}{sensor sampling time [sec]}
\nomenclature{$\mathbf{Q}_{\tau}$}{contaminant tracking matrix}
\nomenclature{$\epsilon_{acc}$}{sensor accuracy provided by manufacturer}
\nomenclature{$\mathbf{Q^*}_{\tau}$}{threshold contaminant tracking matrix}
\nomenclature{$\mathbb{E}$}{expectation operator}
\nomenclature{$\widetilde{\pmb{\xi}}$}{random variable}
\nomenclature{$\rho$}{probability density function}
\nomenclature{$\hat{\rho}$}{approximated probability density function}
\nomenclature{$M$}{number of realizations for random variable $\widetilde{\pmb{\xi}}$}
\nomenclature{$\pmb{\Theta}$}{the weights associated with each sample}

\printnomenclature
\end{document}